\documentclass[paper]{JHEP3}
\usepackage{epsfig}
\usepackage{amssymb}
\def\URLtilde{\lower0.2em\hbox{$\tilde{\phantom{a}}$}}

\def\beq{\begin{equation}}
\def\eeq{\end{equation}}

\def\beqn{\begin{eqnarray}}
\def\eeqn{\end{eqnarray}}
\def\nl{\hfill\break}
\def\s#1{{\small#1}}

\def\HW{\s{HERWIG}}
\def\v{\begingroup\obeyspaces\u}
\def\u#1{\tt#1\endgroup}
\def\ltap{\lower0.2em\hbox{$\stackrel{<}{\sim}$}}

\def\cbar{\hbox{$\bar c$}}
\def\bbar{\hbox{$\bar b$}}
\def\qbar{\hbox{$\bar q$}}

\def\thetac{\hbox{$\Theta_c$}}
\def\thetab{\hbox{$\Theta_b^+$}}

\def\red{
}
\def\black{
}
\def\mycomm#1{\hfill\break\strut\kern-3em{\red\tt ====> #1\black}\hfill\break}
\preprint{
 Cavendish--HEP--04/35\hfill\\
TAUP--2774-04}
\title{\boldmath Coalescence model for $\Theta_c$ pentaquark formation
\footnote{Work supported in part by the UK Particle Physics and
Astronomy Research Council and by the Royal Society.}}
\author{Marek Karliner\\
  Cavendish Laboratory, Madingley Road, Cambridge CB3 0HE, U.K.\\
\kern3em and \\
\hbox{School of Physics and Astronomy,
Raymond and Beverly Sackler Faculty of Exact Sciences}\\
Tel Aviv University, Tel Aviv, Israel\\
  E-mail: \email{marek@proton.tau.ac.il}}
\author{Bryan R.\ Webber\\
  Cavendish Laboratory, 
  Madingley Road, Cambridge CB3 0HE, U.K.\\
  E-mail: \email{webber@hep.phy.cam.ac.uk}}
\abstract{We present a model for the formation of the charmed
pentaquark $\Theta_c$ in hard scattering processes such as deep
inelastic scattering, $e^+e^-$ annihilation, and
 high-energy $p \bar p$ collisions.
The model assumes that the cross section for $\Theta_c$ formation
is proportional to the rate of production of $pD^{*-}$ (or $\bar
pD^{*+}$) pairs in close proximity both in momentum space and in
coordinate space. The constant of proportionality is determined
from the $\Theta_c$ cross section in deep inelastic scattering as
reported by the H1 experiment.  The \HW\ Monte Carlo is used to
generate simulated DIS events and also to model the space-time
structure of the final state.  Requiring the proton and the $D^*$
be within a 100 MeV mass window and separated by a spacelike
distance of no more than 2 fm, we find that a large ``coalescence
enhancement factor'' $F_{\rm co}\sim 10$ is required to account for
the H1 signal.  The same approach is then applied in order to
estimate the number and characteristics of $\Theta_c$ events
produced at LEP and the Tevatron.
  }

\begin{document}

\section{Introduction}

During the last year, many experiments have reported the observation of
an exotic  $uudd\bar s$ pentaquark baryon $\Theta^+$
\cite{Nakano:2003qx,Thetaplus}. Shortly after the initial experimental
reports, Ref.~\cite{Karliner:2003si} postulated the
existence of an anti-charmed analogue \thetac\ with quark content $uudd\cbar$,
and computed its likely properties, using the diquark-triquark
configuration earlier conjectured to describe $\Theta^+$ \cite{OlPenta}.

Ref.~\cite{Karliner:2003si} suggested that
\thetac\ is an isosinglet with $J^P={1\over2}^+$ and estimated its mass at
$2985 \pm 50$ MeV. It also discussed another possible exotic baryon resonance
containing heavy quarks, the \thetab, a $uudd\bar b$ state, and estimated
$M({\thetab})=6398\pm 50$ MeV, noting that
these states should appear as unexpectedly narrow peaks in
$D^-p$, $\bar D^0 n$,
$B^0 p$ and $B^+n$ mass distributions.
Possible search opportunities in $e^+e^-$ have also been suggested in 
Refs.~\cite{Rosner:2003ia,Browder:2004mp,Armstrong:2003zc}.
%\hbox{Refs.~\cite{Rosner:2003ia}-\cite{Armstrong:2003zc}.}

In March, H1 reported \cite{Aktas:2004qf} observing a narrow
resonance in $D^{*-} p$ and $D^{*+} \bar p$ invariant mass
combinations in inelastic $e p$ collisions at $E_{CM}$ of 300 GeV
and 320 GeV at HERA. The resonance has a mass of 3099 $\pm 3$
(stat.) $\pm$ 5 (syst.) MeV and a measured Gaussian width of $12
\pm 3$ (stat.) MeV, compatible with the experimental resolution.
The experiment interpreted this resonance as an anti-charmed
baryon with a minimal constituent quark composition of $uudd\bar
c$, together with the charge conjugate.

A parallel analysis by ZEUS sees no signal \cite{ZEUS_Thetac}.
Null results have also been reported in papers and unpublished reports
at conferences by
FOCUS \cite{FOCUS}, ALEPH \cite{ALEPH} and CDF \cite{CDF}.

Since H1 observes the resonance in both $D^{*-} p$ nd $D^{*+} \bar p$,
it is extremely unlikely that their result is due to a statistical
fluctuation.

Since the ZEUS and H1 experiments are quite similar,
a possible interpretation of these results is that \thetac\ does
not exist and the H1 observation is due to some yet to be understood systematic
effect. 
The discrepancy between H1 and ZEUS
is of a purely experimental nature. It
is now being carefully examined by the two collaborations. 

Putting the ZEUS vs H1 issue aside for the time being, an important
question is: what are the implications of the null results in 
non-HERA experiments?
Since production mechanisms might be different in different experiments,
an {\em a priori} possible interpretation is that these experiments have not
accumulated enough data or that more sophisticated cuts need to be
applied in order to separate the likely small \thetac\ signal from
the background. 
In the present paper we examine the quantitative implications of a possible
H1-like production mechanism at LEP and the Tevatron. 

One can try to apply a simple counting argument to the H1 data
in order to estimate the expected number of events at LEP.
The  H1 data sample of 75 pb$^{-1}$ of DIS events contains roughly
$1.5\times10^6$ events with charm. This should be compared with
the corresponding numbers for $Z^0$ decays at LEP: 
\beqn
\Gamma(\hbox{hadrons})/\Gamma_{\hbox{total}}&=&0.70,
\nonumber\\
\Gamma(c\bar c)/\Gamma(\hbox{hadrons})&=&0.17,
%\\
%\Gamma(b\bar b)/\Gamma(\hbox{hadrons})&=&0.22
%\nonumber
\eeqn
so
$\Gamma(c\bar c)/\Gamma_{\hbox{total}}=0.12$.
%and
%$\Gamma(b\bar b)/\Gamma_{\hbox{total}}=0.15$.
After cuts, each of the LEP experiments has
several million hadronic $Z^0$ decays. Specifically, ALEPH \cite{ALEPH} has 
$3.5\times 10^6$, 
corresponding to approximately
$6\times10^5$ charm events.

We can use this number to make a very crude estimate of the
number of expected anticharm pentaquark events at LEP.
If the production mechanisms at H1 and LEP are roughly similar, the
expected number of anticharmed pentaquarks events at each of the four LEP
experiments is about 40\% of the number of such events at H1.

In the rest of the paper we carry out this estimate in more detail,
with a coalescence model providing a specific possible mechanism for 
$\Theta_c$ formation from $D^* p$,
and HERWIG generating the corresponding $D^*$ and $p$ distributions.

The result of this analysis is that each of the LEP experiments
should have observed between 25 and 40 charmed pentaquark events of the H1
type. This takes into account typical experimental cuts and efficiencies.

At the Tevatron, the integrated luminosity already collected in
Run II is of the order of 200 pb$^{-1}$. If the production mechanism is the
same as in H1, we conclude that even with a very low 
overall detection efficiency of only a few per mille, 
this would provide an unmissable signal 
on the order of $10^5$ events.

In the following sections of the paper we develop the model, extract the
relevant coalescence parameter from the H1 data and then apply it to LEP
and Tevatron experiments. 
We then discuss several possible interpretations of these results.

\section{Coalescence model}
We construct a simple but, we believe, physically appealing model
to attempt a quantitative estimate of the expected
production rate of \thetac\ via coalescence of a $D^*$ and a nucleon.
We use the Monte Carlo program \HW\ \cite{Corcella:2000bw,Corcella:2002jc}
to generate simulated DIS events with HERA kinematics, selecting those
which include either $p$ and $D^{*-}$ or $\bar p$ and $D^{*+}$.  In
addition to the momentum-space structure of final states, \HW\ generates
an approximate space-time structure, which enables us to
model the joint distribution of invariant mass and space-time separation
of the $pD^*$ pairs.  Assuming resonance formation through coalescence,
we suppose that the cross section for \thetac\ production is proportional
to the number of $pD^*$ pairs that have invariant masses in a small
interval $\Delta m$ around the resonance mass and are at the same time
formed within a small spacelike separation  $\Delta x$.
By adjusting the constant of proportionality to match the \thetac\ signal
observed in DIS by H1, we predict the \thetac\ signal that should be
seen in other processes with a similar production mechanism.

Before going into a detailed discussion of the model's implementation, 
a word of warning is in order. 
The present work presents a new type of approach, 
going beyond the standard coalescence model. This is made possible by the 
fact that HERWIG
generates a space-time distribution of hadrons produced
in high-energy collisions. As with any new approach, one would like to
first test it against other similar cases where the answer is known.
This can be done to some extent by looking at (anti)-deuteron production 
at HERA and at LEP, as discussed at the end of Sec. 2.3.
But the experimental situation with deuterons is puzzling on its own, and
the analogy can be only pushed so far. As for other well-known resonances,
their dominant production
mechanism is hadronization of quark and gluon jets, rather than through
coalescence. The relevant
yields of such resonances are built into HERWIG from the beginning
and closely follow the experimental data.
However, the charmed pentaquark, and especially the baffling
experimental situation, is sufficiently interesting that we believe it
provides strong motivation for the study of the production mechanism,
even though
the model cannot be directly tested on other well-known resonances.

\subsection{Model for space-time structure}
Since the space-time structure of the final state is a crucial ingredient
of the proposed coalescence mechanism, we start with a brief summary of the
\HW\ model for this structure.  Fuller details may be found in
ref.~\cite{Corcella:2000bw}.

During development of the final state, \HW\ assigns to each
unstable object with four-momentum $q$ an exponentially
distributed proper lifetime, with mean value $\tau$ given
by~\cite{Sjostrand:1993hi}
\beq
\tau =\frac{\hbar\sqrt{q^2}}{\sqrt{(q^2-M^2)^2 + (\Gamma q^2/M)^2}}
\eeq
where $M$ and $\Gamma$ are the nominal mass and width.
This formula interpolates between $\tau=\hbar/\Gamma$ for an object
that is on mass-shell and $\tau=\hbar\sqrt{q^2}/(q^2-M^2)$ for one that
is far off mass-shell.  Following sequential electroweak decays and QCD parton
showers, the resulting space-time separations are combined to give the
locations of partons at the start of the hadronization process.

For light quarks and gluons, whose natural widths are small, the above
prescription can lead to unreasonably large distances being generated in the
final, low-virtuality steps of parton showering. To avoid this they are given
a width $\Gamma=\v{VMIN2}/M$, the \HW\ parameter \v{VMIN2} (default value
0.1~GeV$^2$) acting as a lower limit on a parton's virtuality.

Hadronization in \HW\ takes place through non-perturbative gluon splitting,
$g\to q\qbar$, followed by colour-singlet $q\qbar$ cluster formation and
decay. The production vertex of a cluster is taken as the midpoint of a line
perpendicular to the cluster's direction of travel, with its two ends
on the trajectories of the constituent quark and antiquark.
The production positions of primary hadrons from cluster decays are
smeared, relative to the cluster position, according to a Gaussian
distribution of width $\hbar/M$ where $M$ is the mass of the cluster.

\subsection{DIS results}
We first use \HW\ to simulate DIS in $e^+p$ collisions with beam momenta
$27.6\times 920$ GeV/c and integrated luminosity 75 pb$^{-1}$, which
approximates the data sample used in the H1 analysis~\cite{Aktas:2004qf}.
For greater efficiency, we limit the target partons to charmed quarks
and antiquarks only, by setting the process code $\v{IPROC}=9004$.
Using the \HW\ default (MRST LO~\cite{Martin:1998np})
parton distribution functions, we find a
total leading-order DIS-on-charm cross section of $18.4$ nb for momentum
transfers in the region $6<Q^2<100$ GeV$^2$ and inelasticities
$0.05<y<0.7$.  We therefore generate $1.4 \times 10^6$ simulated
events so as to correspond roughly to the 75 pb$^{-1}$ integrated
luminosity of the H1 data sample.  In \HW\ runs with $\v{IPROC}=9000$
(DIS on all flavours of partons), we find that the contribution to $pD^*$
production in the region of interest from DIS on non-charmed partons
is negligible.

We use a momentum transfer range of $6<Q^2<100$ GeV$^2$,
rather than the full range $1<Q^2<100$ GeV$^2$ covered by the H1 analysis,
because the charm PDF and \HW\ simulation are not reliable at lower $Q^2$.
The predicted dependence of the $D^{*\pm}$ cross section on $Q^2$ is shown
in fig.~\ref{fig:Dstar_Q2} together with the recent H1 data~\cite{H1_charm}.
The predicted $D^*$ pseudorapidity distribution is shown
with the H1 data in fig.~\ref{fig:Dstar_eta}.
These results can also be compared with 
ZEUS results~\cite{ZEUS_charm} for charm production in DIS.
 
\FIGURE{
\epsfig{figure=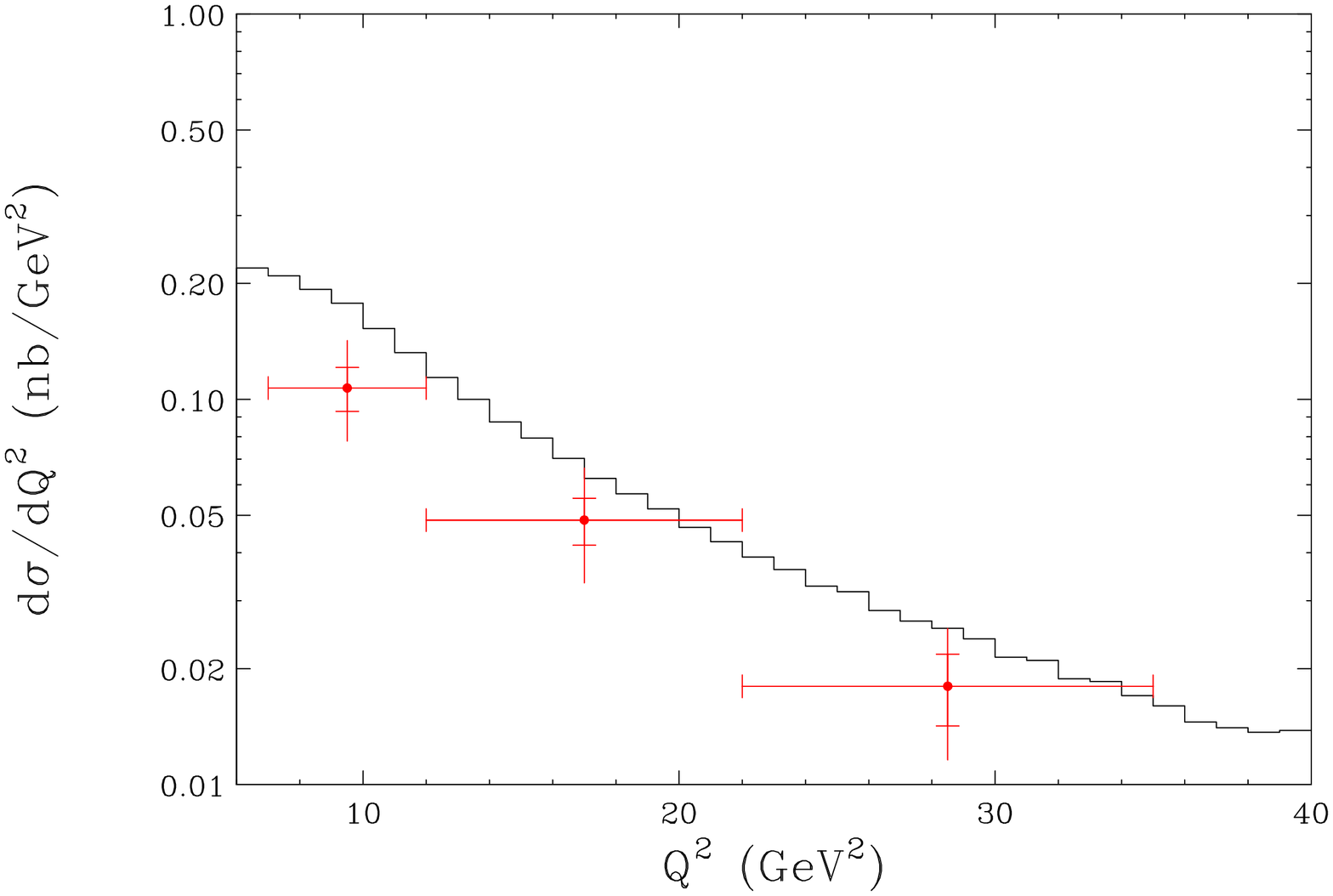,angle=90,width=0.8\textwidth}
\caption{\HW\ prediction for $D^{*\pm}$ production in DIS at $Q^2>6$ GeV$^2$,
together with recent H1 data \cite{H1_charm}.
\label{fig:Dstar_Q2}}
}
\TABLE{
\begin{tabular}{|c|c|}\hline
$p_t(p)_{\rm min}$ & 120 MeV \\
$p_t(D^*)_{\rm min}$ & 1.5 GeV \\
$\eta(D^*)_{\rm min}$ & --1.5 \\
$\eta(D^*)_{\rm max}$ & 1.0 \\
$z(D^*)_{\rm min}$ & 0.2 \\
\hline
\end{tabular}
\caption{Selection criteria
 \break
 for $pD^*$ production.}
\label{tab:prod_cuts} }

\FIGURE{
\epsfig{figure=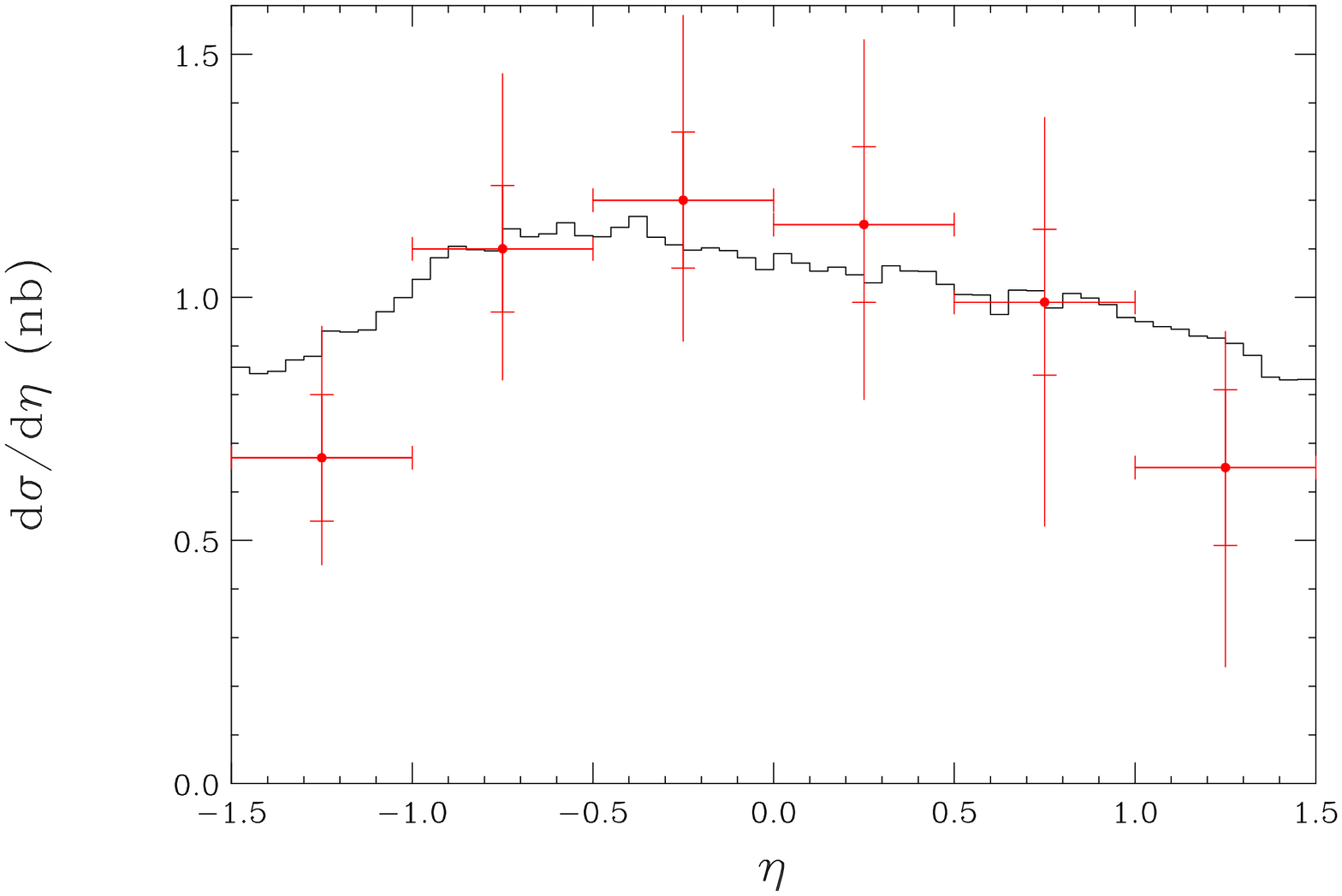,angle=90,width=0.8\textwidth}
\caption{Predicted $D^{*\pm}$ pseudorapidity distribution in DIS, together
with recent H1 data \cite{H1_charm}.
\label{fig:Dstar_eta}}
}

The histograms in figs.~\ref{fig:Dstar_Q2} and \ref{fig:Dstar_eta} 
are for the cuts in ref.~\cite{Aktas:2004qf}, while the data
shown are for slightly different cuts (see ref.~\cite{H1_charm} for details), 
which we estimate may reduce the cross section by 30-40\% without changing the
shape significantly. In view of the more limited range of $Q^2$ accessible
in the \HW\ simulation, a detailed comparison with the data is not possible
at present, but we judge the semi-quantitative agreement to be satisfactory.

Next we select events in which both a
proton and a $D^{*-}$, or an antiproton and a $D^{*+}$, are
produced.  Imposing the cuts in Table~\ref{tab:prod_cuts}, which
approximate those applied in the H1 analysis, but not yet
selecting particular $D^*$ decays, we obtain the $pD^*$ mass
spectrum shown by the solid histogram in fig.~\ref{fig:m_dist}.
%\eject
\FIGURE{
\epsfig{figure=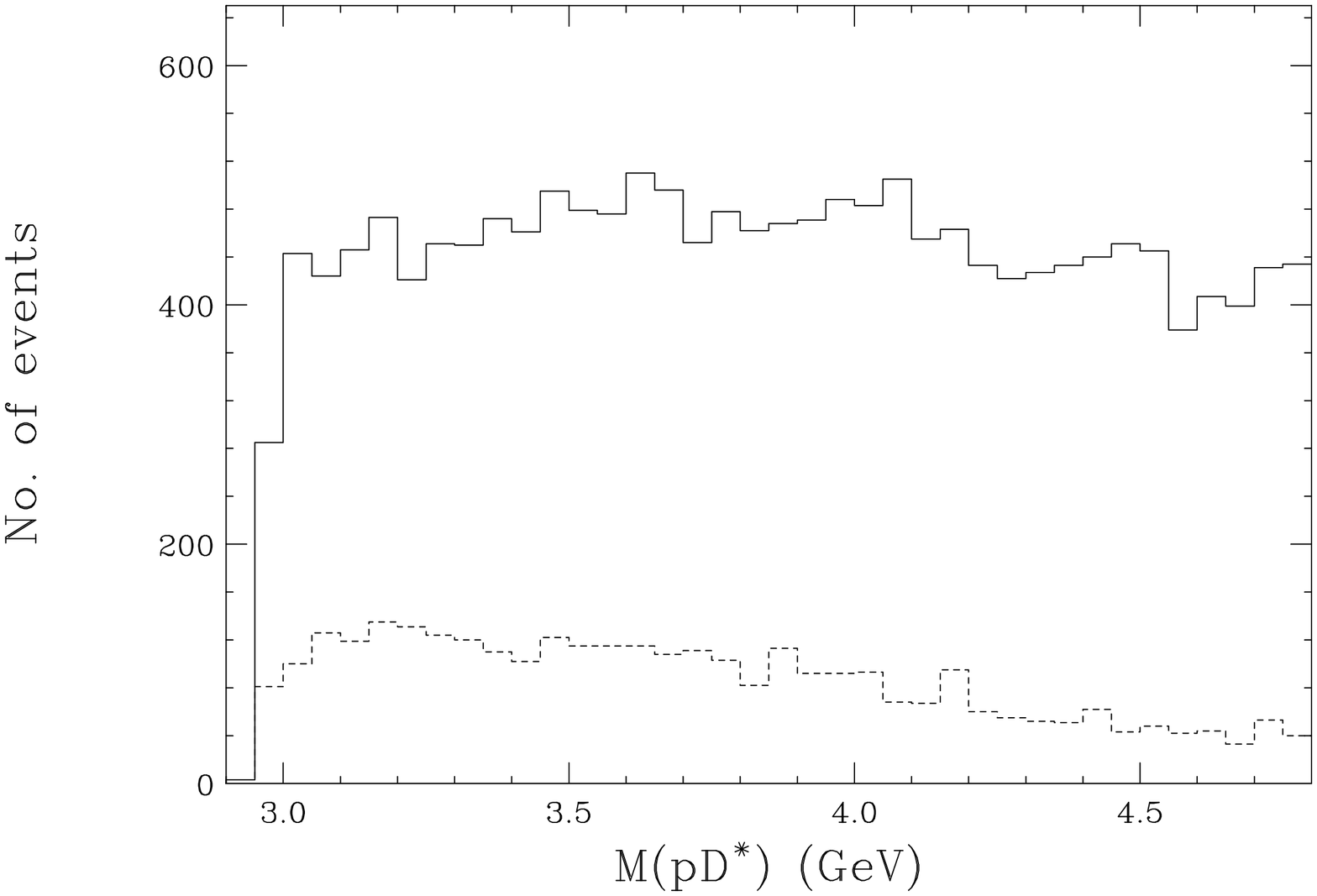,angle=90,width=0.8\textwidth} \caption{Predicted
non-resonant $pD^*$ mass distribution in DIS. Solid: all $pD^{*-}$
and $\bar pD^{*+}$ pairs. Dashed: pairs with separation $0<\Delta
x<2$ fm. \label{fig:m_dist}} }
%\eject

\FIGURE{
\epsfig{figure=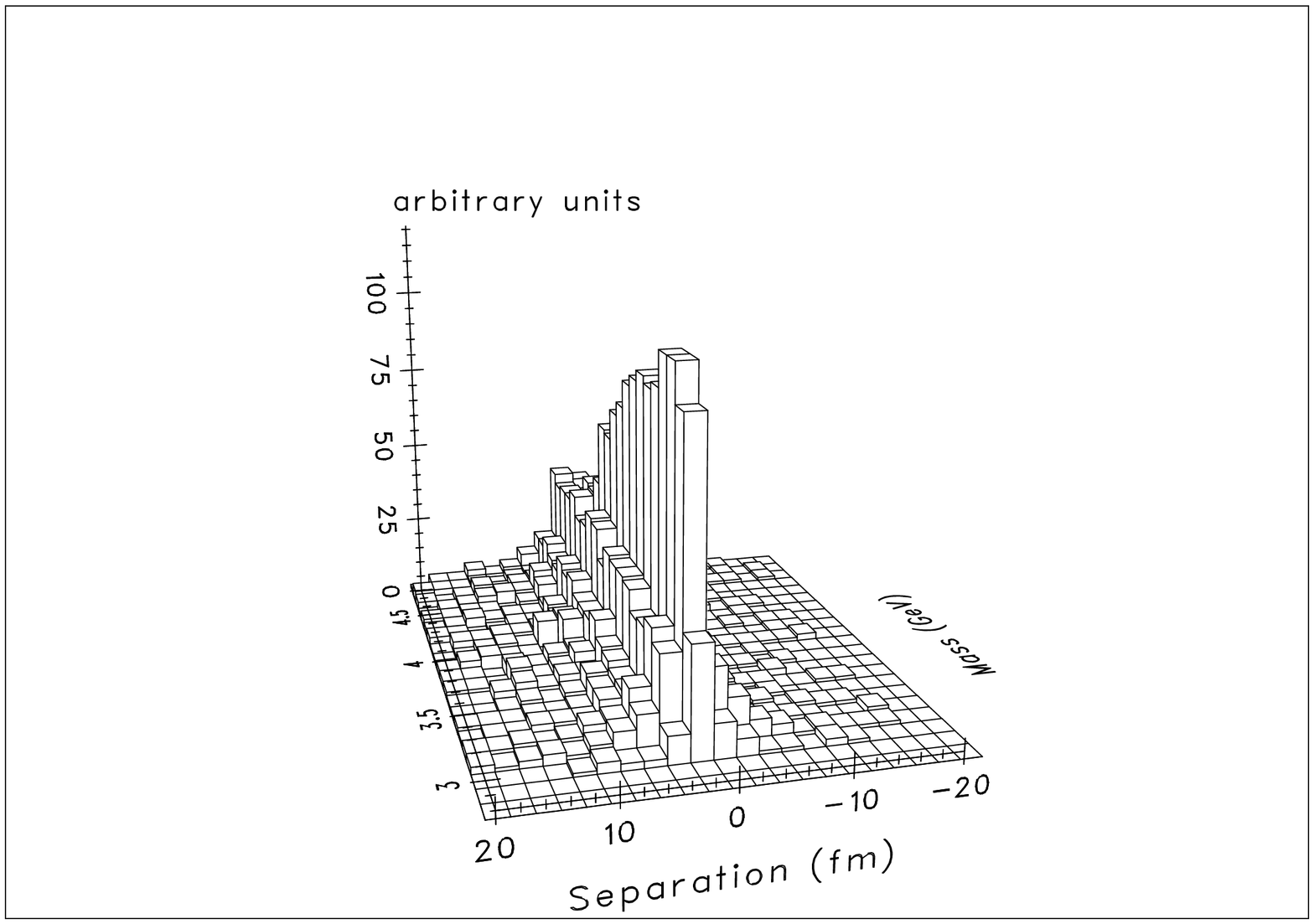,angle=90,width=0.96\textwidth}
\caption{Predicted $pD^*$ joint mass--separation distribution in
DIS. \label{fig:m_sep_dist}} }
The predicted joint distribution in mass and space-time separation
is shown in fig.~\ref{fig:m_sep_dist}. Here we depict spacelike
separations as positive and timelike as negative.  There is a
preference for spacelike separations, peaking at small values,
corresponding to the formation of primary hadrons on a hyperboloid
at a roughly universal proper hadronization time.  This should be
a general characteristic of physically reasonable hadronization
models. 

 We now suppose that the \thetac\ resonance, not
explicitly present in the \HW\ hadronization scheme, can be
modelled as an enhancement in the cross section proportional to
the number of $pD^*$ pairs with invariant masses in a window
$\Delta m$ close to resonance and at the same time within the
range  $\Delta x$ of the binding mechanism.  The measures of
closeness should be on the hadronic scale of $\Delta m\sim 100$
MeV, $\Delta x\sim 1$ fm. To be specific, we choose the mass
interval $3050<m<3150$ MeV and spacelike separations $0<\Delta
x<2$ fm.

The predicted mass distribution for $pD^*$ pairs with separations
$0<\Delta x<2$ fm is shown by the dashed histogram in fig.~\ref{fig:m_dist}.
Of these pairs, 245 fall within the coalescence region $3050<m<3150$ MeV.
With a coalescence enhancement factor $F_{\rm co}$, we therefore expect
of the order of $250 F_{\rm co}$ events with \thetac\ production.

\TABLE{
\begin{tabular}{|c|c|}\hline
$p_t(\pi_s)_{\rm min}$ & 120 MeV \\
$p_t(K)_{\rm min}$ & 500 MeV \\
$p_t(\pi)_{\rm min}$ & 250 MeV \\
$[p_t(K)+p_t(\pi)]_{\rm min}$ & 2 GeV\\
\hline
\end{tabular}
\caption{Selection criteria for
\break
 $D^*\to D^0\,\pi_s$,
$D^0\to K^-\pi^+$ decay.} \label{tab:dec_cuts} } 

In the H1
experiment, the $D^{*\pm}$ is detected through the decay sequence
$D^{*\pm}\to D^0\pi^\pm$, $D^0\to K^-\pi^+$, which has a branching
fraction of about 3\%. Taking into account the H1 cuts needed to
identify the decay products, listed in Table~\ref{tab:dec_cuts},
we estimate that this fraction 
is reduced to roughly 2\%.
Therefore the expected H1 \thetac\ signal is about $5 F_{\rm co}$
events. Given the observed signal of 50 events, this implies a
rather large
coalescence enhancement factor of $F_{\rm co}\simeq 10$.

To obtain the observed $pD^*$ mass distribution from the one
shown by the solid histogram in fig.~\ref{fig:m_dist}, one should
multiply by the $D^{*\pm}$ detection efficiency.  The shape of the
distribution is similar to that shown in ref.~\cite{Aktas:2004qf},
but the above efficiency estimate of 2\% implies an observed number
of about 10 pairs per 50 MeV bin, which is considerably smaller
than the number of about 10 pairs per 10 MeV bin found by H1.
This could be due to an underestimate of the proton yield by \HW,
a higher relative proton yield in charm production with
$1<Q^2<6$ GeV$^2$, or a large contamination of the
experimentally reported distribution by non-protons.

We expect the predicted proton yield in the
coalescence region to be more reliable, since additional soft
contributions will tend to be more widely distributed in space-time.
We therefore use the value $F_{\rm co}\simeq 10$ throughout the
rest of this paper.
However, it should be borne in mind that an enhancement of the
proton density in the coalescence region in DIS would imply a reduced
value of the coalescence factor $F_{\rm co}$ and hence a
smaller prediction of \thetac\ production in other processes.
On the basis of the information available to us, we estimate
the maximal range of values indicated by the H1 signal to be
$2\ltap F_{\rm co}\ltap 10$.   

\FIGURE{
\epsfig{figure=thetac_Q2.ps,width=0.75\textwidth}
\caption{Predicted $\Theta_c$ production cross section as a function of
momentum transfer in DIS.\label{fig:thetac_Q2}}
}

Of course, changing the size of the coalescence region by varying
$\Delta m$ and/or $\Delta x$ can be compensated by a change in
$F_{\rm co}$ to give the same signal.  For example, one could
increase the mass interval to 200 MeV, i.e. $3000<m<3200$ MeV, and
the maximum separation to 4 fm.  Then a value of $F_{\rm co}\simeq 2.5$
would account for the H1 signal, without changing substantially
the predictions given below. This stability against variation of
$\Delta m$ and/or $\Delta x$ provides a nontrivial test of the
coalescence model's robustness.

Having fixed the coalescence factor $F_{\rm co}$, we can make
predictions about other aspects of \thetac\ production in DIS.
For example, we show in fig.~\ref{fig:thetac_Q2} the predicted
\thetac\ production cross section in DIS as a function of the
momentum transfer $Q^2$.  The rather weak dependence at
low $Q^2$ helps to justify our exclusion of the region
$Q^2<6$ GeV$^2$.
Figures \ref{fig:thetac_eta} and \ref{fig:thetac_pt} show the predicted
pseudorapidity and transverse momentum distributions of the $\Theta_c$,
computed from the corresponding
quantities for $pD^*$ pairs in the coalescence region.

\FIGURE{
\epsfig{figure=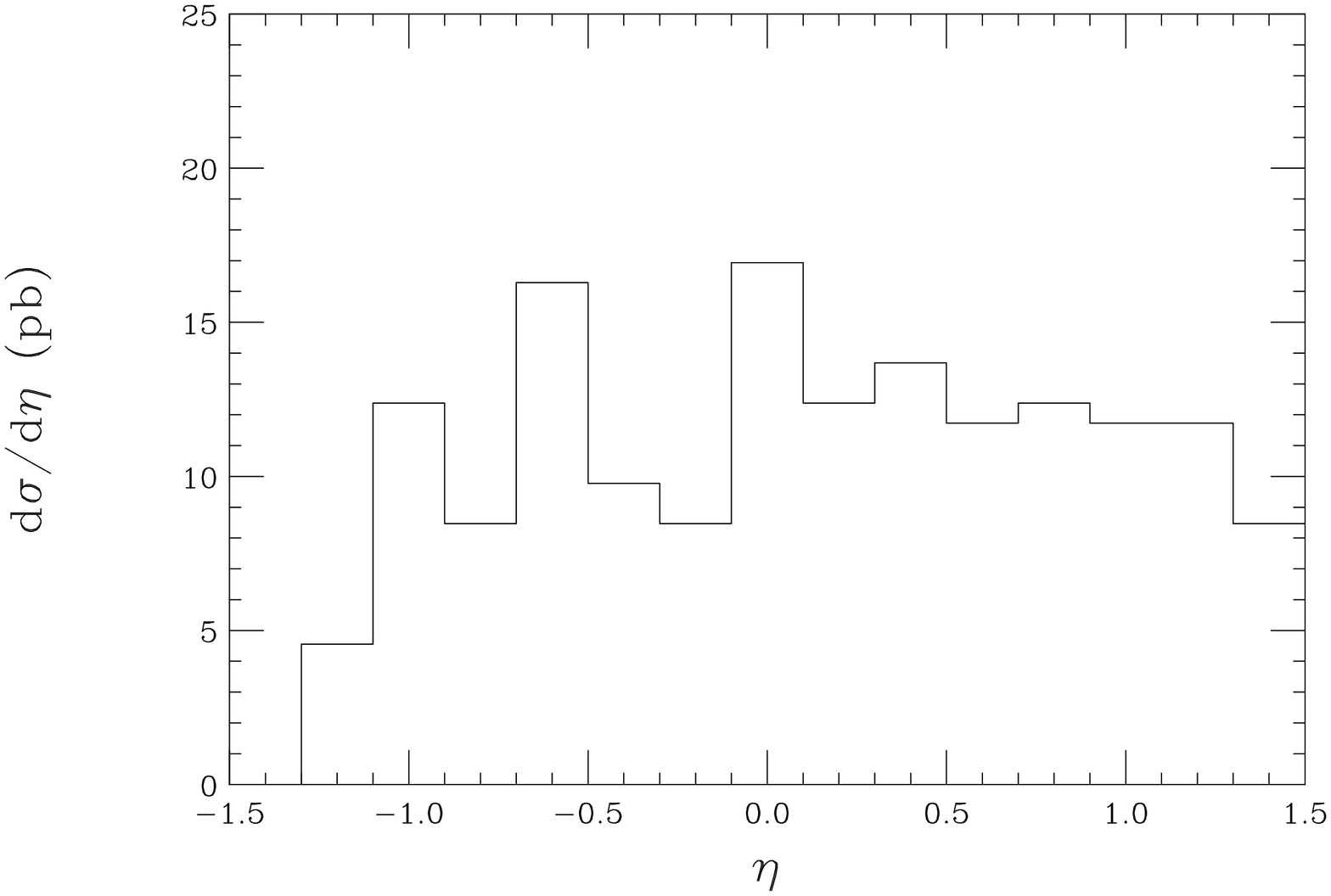,angle=90,width=0.75\textwidth}
\caption{Predicted $\Theta_c$ pseudorapidity distribution in DIS.
\label{fig:thetac_eta}}
}
\FIGURE{
\epsfig{figure=thetac_pt.ps,width=0.75\textwidth}
\caption{Predicted $\Theta_c$ transverse momentum distribution in DIS.
\label{fig:thetac_pt}}
}

\eject
\subsection[$e^+e^-$ results]{\boldmath $e^+e^-$ results}
The coalescence model can be used to predict the rates and
distributions of \thetac\ production in other hard processes
that can be generated by \HW, such as $e^+e^-$ annihilation.
We simply assume that the \thetac\ distributions are always
in the fixed ratio $F_{\rm co}=10$ to those of $pD^*$ pairs
in the coalescence region of $\Delta m$ and $\Delta x$.

We concentrate on the resonant reaction $e^+e^-\to Z^0\to$ hadrons
since this is the place where the LEP experiments accumulated most
of the data that might show evidence of \thetac\ production. We
generate $7\times 10^5$ simulated events of the process $e^+e^-\to
Z^0\to c\cbar$ at the resonance peak, which corresponds to the
approximate integrated luminosity of 140 pb$^{-1}$ collected by
each experiment at or near $Z^0$ resonance.  Although the process
$e^+e^-\to Z^0\to b\bbar$ is a copious indirect source of charm,
the possible production of \thetac\ in $b$-decay is a separate
issue and we assume that this source could be distinguished
experimentally from direct production.  As in the case of DIS, we
find that, within the framework of the \HW\ model, the
contribution to $pD^*$ production in the coalescence region from
light quark production ($e^+e^-\to Z^0\to q\qbar$ where $q=u,d,s$)
is negligible.

We apply no selection criteria like those in
Tables~\ref{tab:prod_cuts} and \ref{tab:dec_cuts}, since these are
mostly irrelevant to the LEP environment.  Therefore our
predictions represent upper limits on the observable signal at
LEP1 for consistency with the H1 data and the coalescence model.

Figure~\ref{fig:m_sep_dist_zdk} shows the resulting $pD^*$ joint
distribution in mass and space-time separation, to be compared with
fig.~\ref{fig:m_sep_dist} for DIS. The two distributions are similar,
with some additional narrowing of the mass distribution at small
separations in the case of $Z^0$ decay.
\FIGURE{
\epsfig{figure=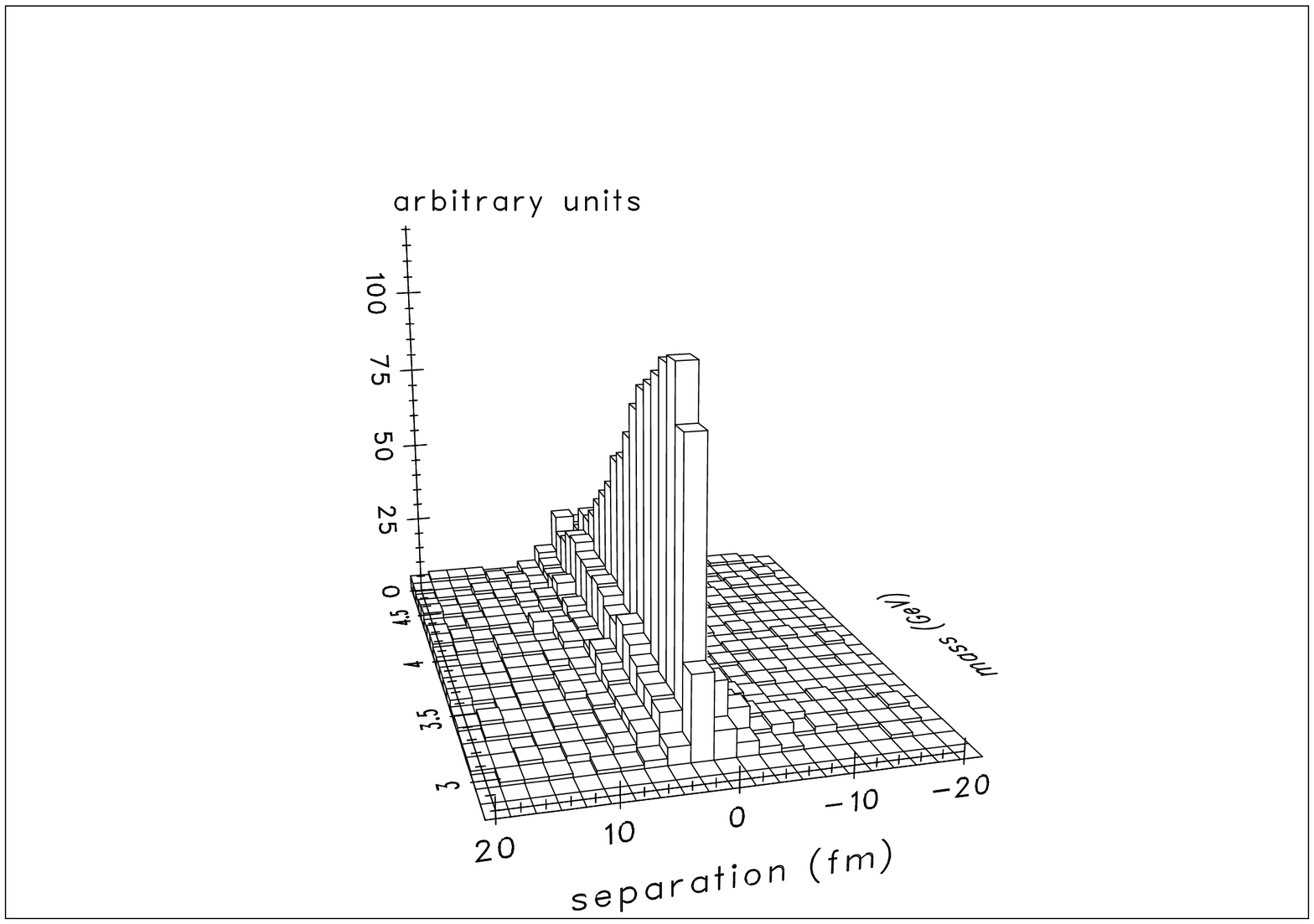,angle=90,width=0.93\textwidth}
\caption{Predicted $pD^*$ joint mass--separation distribution in
$Z^0$ decay.
\label{fig:m_sep_dist_zdk}}
}
\FIGURE{
\epsfig{figure=m_dist_zdk.ps,width=0.8\textwidth}
\caption{Predicted non-resonant $pD^*$ mass distribution in $Z^0$ decay.
Solid: all $pD^{*-}$ and $\bar pD^{*+}$ pairs. Dashed: pairs with
separation $0<\Delta x<2$ fm.
\label{fig:m_dist_zdk}}
}

The overall $pD^*$ mass distribution is shown by the solid histogram
in fig.~\ref{fig:m_dist_zdk}, corresponding to
fig.~\ref{fig:m_dist} for DIS.
In order to make a comparison with experiment, we need to take into
account data reduction due to cuts and efficiencies. 
We generated $7\times 10^5$ charm events, corresponding to 
$7\times 10^5/0.17 = 4.1\times 10^6$ hadronic events, to be compared with
$3.5\times10^6$ hadronic events reported by ALEPH \cite{ALEPH} 
after cuts. ALEPH reports 1.3M proton candidates, whereas the proton yield
is about 1 per event, so their mean proton efficiency is $\sim 1/3$.
For the D* efficiency, through
$D^{*\pm}\to D^0\pi^\pm$, $D^0\to K^-\pi^+$,
we assume a range 2-3\%.
These cut and efficiencies yield a reduction factor between 
$5.7\times10^{-3}$ and $8.5\times10^{-3}$. 

\eject

In \cite{ALEPH} ALEPH reports 82 $D^{*-}p$ events in the mass range 
$2.95~\hbox{GeV} < M(D^{*-}p) < 4~\hbox{GeV}$. 
The total number of $D^{-*}p$ events in this mass range in
fig.~\ref{fig:m_dist_zdk} is about
15,000, so we would expect between 85 and 130 events to be seen, in good
agreement with \cite{ALEPH}.

The mass distribution after a cut on separation is
shown by the dashed histogram in fig.~\ref{fig:m_dist_zdk}.
The number of pairs in the
coalescence region $0<\Delta x<2$ fm, $3050<m<3150$ MeV is 576,
roughly as expected from the 245 found in DIS with about
half the integrated luminosity.
The corresponding \thetac\ production cross section
on resonance, assuming a coalescence enhancement factor of
$F_{\rm co}\simeq 10$, is 40 pb. 
Taking into account the data reduction factors above, that
implies a signal of about 25-40 events per experiment in the
observed decay channel.

The predicted distribution of \thetac\ production with respect to
the momentum fraction $x_p=2|{\bf p}|/\sqrt{s}$ is shown in
fig.~\ref{fig:thetac_xp_zdk}.  This distribution is similar in shape
to that of the $D^{*\pm}$, but slightly narrower
(fig.~\ref{fig:Dstar_xp_zdk}).
\FIGURE{
\epsfig{figure=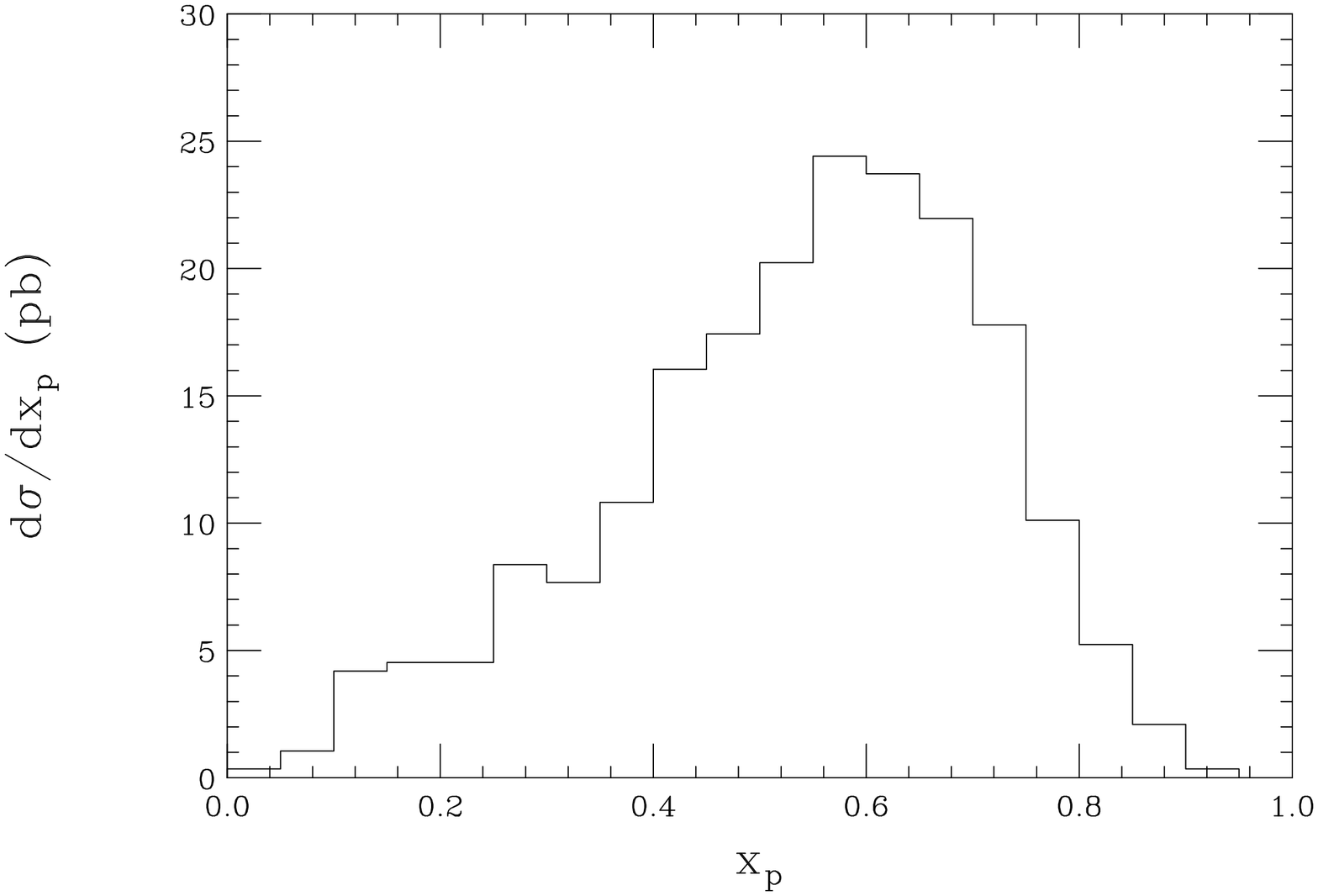,angle=90,width=0.8\textwidth}
\caption{Predicted cross section $e^+e^-\to Z^0\to\thetac X$
as a function of momentum fraction.\label{fig:thetac_xp_zdk}}
}
\FIGURE{
\epsfig{figure=Dstar_xp_zdk.ps,width=0.8\textwidth}
\caption{Predicted  cross section $e^+e^-\to Z^0\to D^{*\pm}X$
as a function of momentum fraction.\label{fig:Dstar_xp_zdk}}
}
The predicted \thetac\ production is strongly collimated along the jet axes of
the final state, as shown by the pseudorapidity and transverse momentum
distributions relative to the thrust axis
(figs.~\ref{fig:thetac_eta_zdk}, \ref{fig:thetac_pt_zdk}).
\eject
\FIGURE{
\epsfig{figure=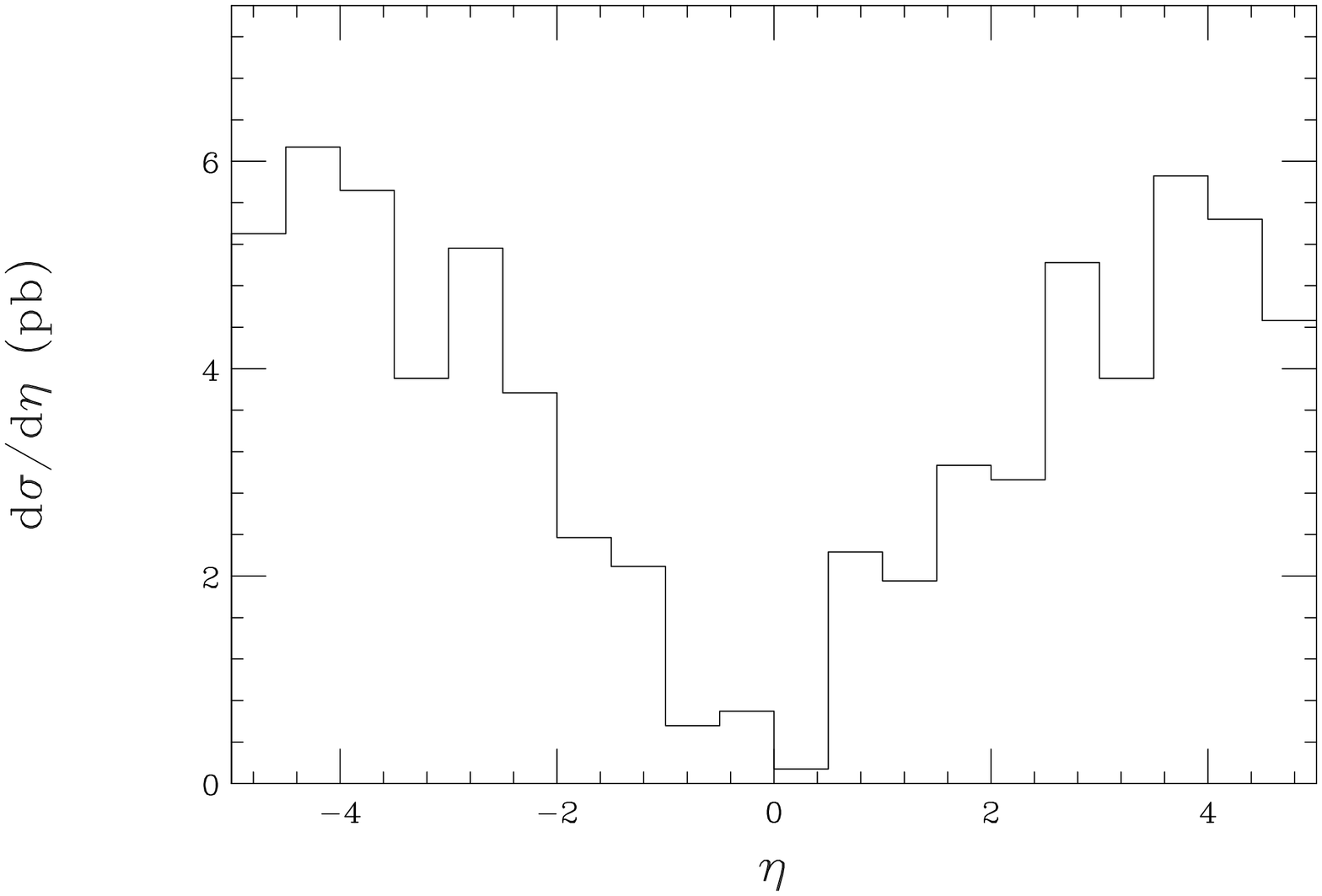,angle=90,width=0.8\textwidth}
\caption{Predicted cross section $e^+e^-\to Z^0\to\thetac X$ as a function of
 pseudorapidity relative to the thrust axis.\label{fig:thetac_eta_zdk}}
} \FIGURE{ \epsfig{figure=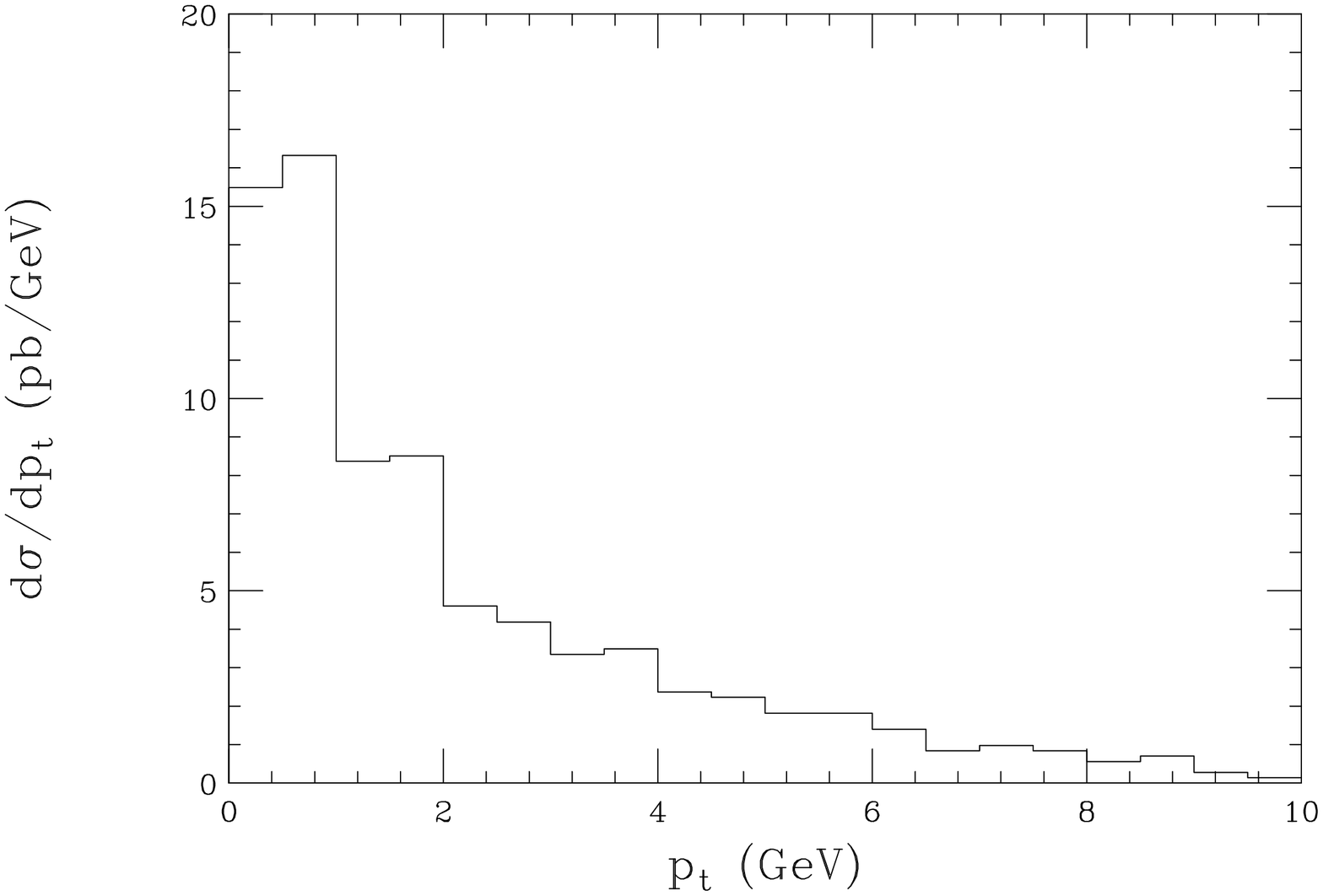,angle=90,width=0.8\textwidth}
\caption{Predicted cross section $e^+e^-\to Z^0\to\thetac X$ as a
function of transverse momentum relative to the thrust
axis.\label{fig:thetac_pt_zdk}} }

% the \eject and \strut commands are needed for proper 
% placement of text and figures
\eject
\strut
\eject
\strut
\eject

In contrast,
ALEPH reports \cite{ALEPH} no deviation from background in the relevant
mass region, i.e. no signal.
Our results then have several possible interpretations:
\begin{itemize}
\item[(a)]
there is no resonance and the H1 signal is due to some yet to be understood
systematic effect;
\item[(b)]
H1 observed a genuine resonance, but for some reason 
the production mechanism in DIS is different than in $e^+e^-$;
\item[(c)]
the production mechanism is the same in DIS and $e^+e^-$
but it is not described by the coalescence model;
\item[(d)]
some crucial ingredient in the $e^+e^-$ data
analysis is not well understood.
\end{itemize}

In this context
we would like to draw the reader's attention to two examples of 
non-exotic baryon states which are definitely known to exist, but whose 
study seems to be problematic at LEP.

The first example is $\Lambda_c(2880)^+$ which is above the $N D$
threshold.
According to RPP \cite{RPP}, the only observed decay modes of
$\Lambda_c(2880)^+$ are $\Lambda_c^+ \pi^+ \pi^-$ and $\Sigma_c(2550) \pi$.
So is unclear why no one has seen $\Lambda_c(2880)^+ \rightarrow DN$
($D^*N$ is above threshold).

An even more striking example is the non-observation of (anti)-deuterons at LEP.
As already noted elsewhere \cite{Karliner:2004gr},
it is also useful to compare the rate of
antideuteron and antiproton production in a given experiment.
Such an analysis has been carried out by H1 \cite{Aktas:2004pq},
yielding an antideuteron/antiproton ratio of \ $\bar d/\bar p = 5.0 \pm
1.0 \pm 0.5\times10^{-4}$ in photoproduction.

On the other hand, although the LEP experiments produced roughly one
proton per $Z^0$ decay \cite{Knowles:1997dk} and have accumulated
millions of $Z^0$ decays on tape, very little is known about
antideuteron production at LEP. The one theoretical prediction we
are aware of is Ref.~\cite{Gustafson:1993mm}, which uses the Lund
string fragmentation model to predict  $5 \times 10^{-5}$
deuterons per $Z^0$ decay. The only relevant experimental
publication we are aware of is from OPAL \cite{Akers:1995az},
which reports exactly {\em one} antideuteron candidate event which
was eventually dismissed because it did not pass through the
primary vertex. From this OPAL infers at 90\% confidence level an
upper limit on antideuteron production of $0.8 \times 10^{-5}$
antideuterons per $Z^0$ in the momentum range $0.35 < p < 1.1$
GeV.

A recent estimate \cite{SloanPC} based on this data concludes that
$\bar d/\bar p < 1.6 \times 10^{-4}$, which is significantly less
than the ratio reported by H1  \cite{Aktas:2004pq}.\footnote{We
thank T.~Sloan for discussion of this point.} The reason for this
presumed difference is unknown at present.
In analogy with $\Theta_c$ formation, 
one can construct a coalescence model for deuteron formation.
Using a similar coalescence window  
\hbox{($m-m_p-m_n<100$~MeV,} $0<\Delta x<2$ fm), 
the LEP upper limit implies a coalescence factor for $\bar p+\bar n\to \bar d$,
of $F_{\rm co}^{(d)} < 0.1$.
It would be very
valuable to have more information on antideuterons from the LEP
experiments.

%\strut\eject
\subsection{Hadroproduction results}
The \thetac\ coalescence model 
can also be used to predict its production in hadron-hadron
collisions.  The only limitation, as in DIS, is that \HW\ is able to
model only hard processes.\footnote{The results in
this section include correction of an error in HERWIG 
in the programming of the
space-time structure of the underlying event in hadron-hadron collisions.
The quantitative changes are small.}
 In heavy quark hadroproduction the quark mass
provides a hard scale, but in the case of charm this is not large enough to
ensure perturbative reliability.  We therefore set a minimum transverse
momentum for the hard subprocess, setting the \HW\ parameter {\tt PTMIN} 
to 5 GeV.  This means that charmed hadron production can only be predicted
at transverse momenta greater than this value.

The prediction for $D^{*\pm}$ production above 5 GeV in Tevatron Run II
($p\bar p$ at $\sqrt{S}=2$ TeV), shown in fig.~\ref{fig:Dstar_pt_tev}, is
indeed in reasonable agreement with CDF preliminary data~\cite{Bussey:2004xf}.
\FIGURE{
\epsfig{figure=Dstar_pt_tev.ps,width=0.8\textwidth}
\caption{\HW\ prediction for $D^{*\pm}$ transverse momentum distribution
in $p\bar p$ collisions at $\sqrt{S} = 2$ TeV.
\label{fig:Dstar_pt_tev}}
}

The model study presented here is based on a sample of $10^5$
simulated charm production events. The \HW\ estimate of the charm
cross section with the above transverse momentum cut is 0.15 mb.
Therefore the equivalent integrated luminosity of our sample is
quite small, around 0.7 nb$^{-1}$.

Figure~\ref{fig:m_sep_dist_tev} shows the corresponding $pD^*$ joint
distribution in mass and space-time separation, to be compared with
figs.~\ref{fig:m_sep_dist} and \ref{fig:m_sep_dist_zdk} for DIS and
$Z^0$ decay, respectively.  We see that the distribution in hadroproduction
is substantially broader in both variables.  This is due to the more
complex structure of hadron-hadron collisions.  In particular, the
underlying event (due to interactions between the beam hadron remnants)
often contains protons and antiprotons formed over a longer timescale
than that required for coalescence.
\FIGURE{
\epsfig{figure=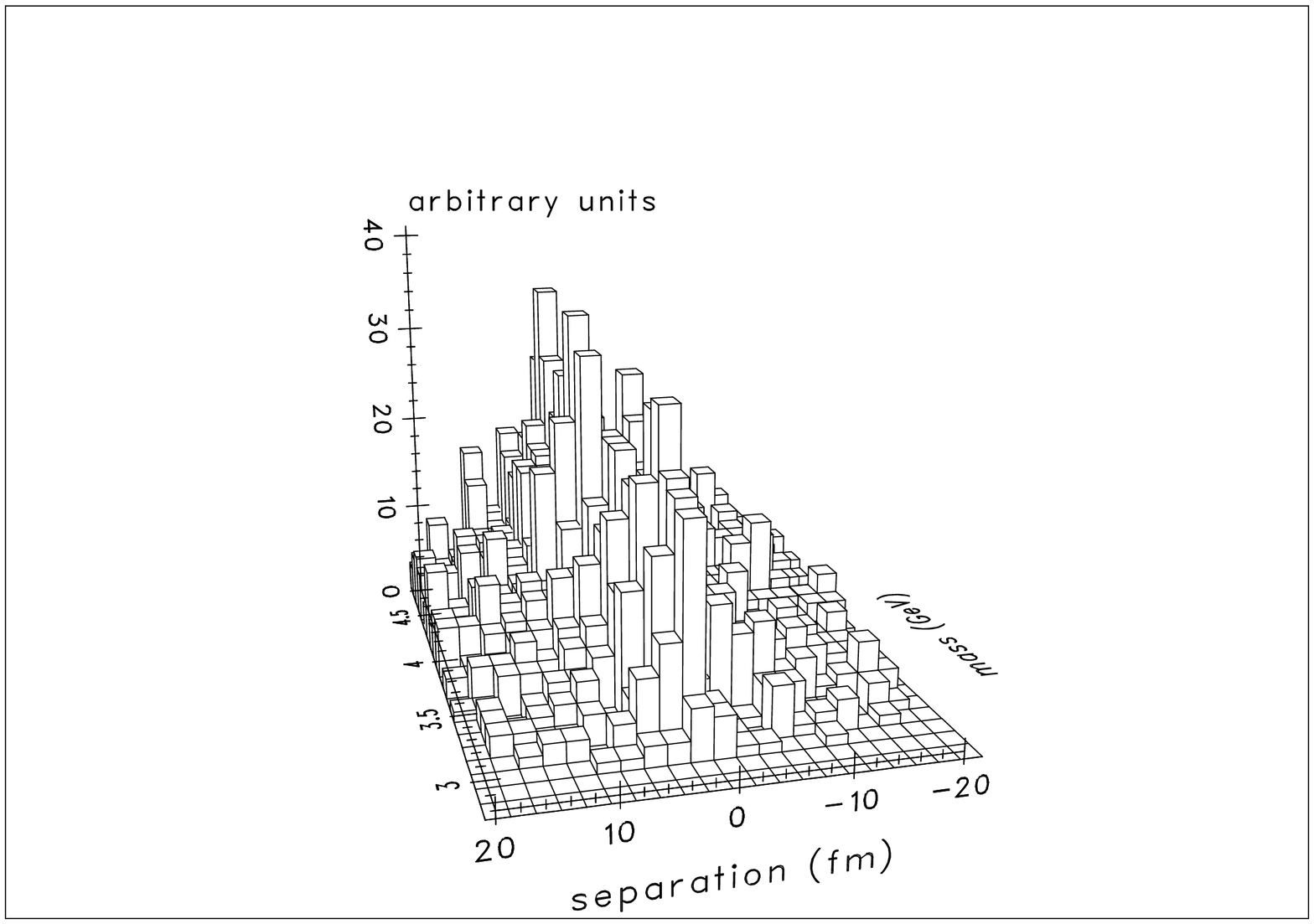,angle=90,width=0.85\textwidth}
\caption{Predicted $pD^*$ joint mass--separation distribution in
charm hadroproduction at $\sqrt{S} = 2$ TeV.
\label{fig:m_sep_dist_tev}}
}
\FIGURE{
\strut\kern-1em
\epsfig{figure=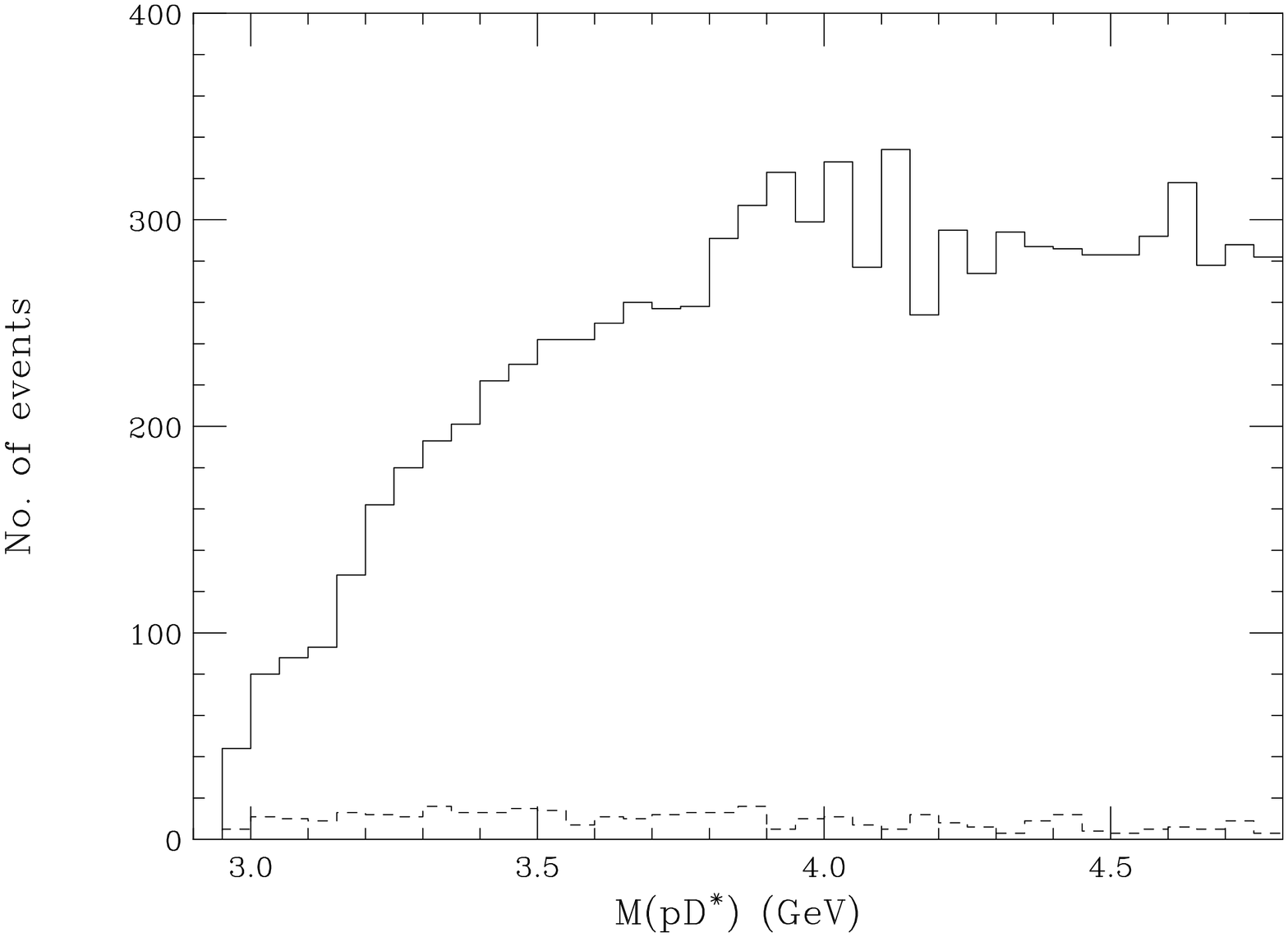,width=0.57\textwidth,angle=90}
\caption{Predicted non-resonant $pD^*$ mass distribution in charm
hadroproduction at $\sqrt{S} = 2$ TeV.
Solid: all $pD^{*-}$ and $\bar pD^{*+}$ pairs. Dashed: pairs with
separation $0<\Delta x<2$ fm.
\label{fig:m_dist_tev}}
}

%\strut
%\eject
%\strut
%\eject

\FIGURE{
\epsfig{figure=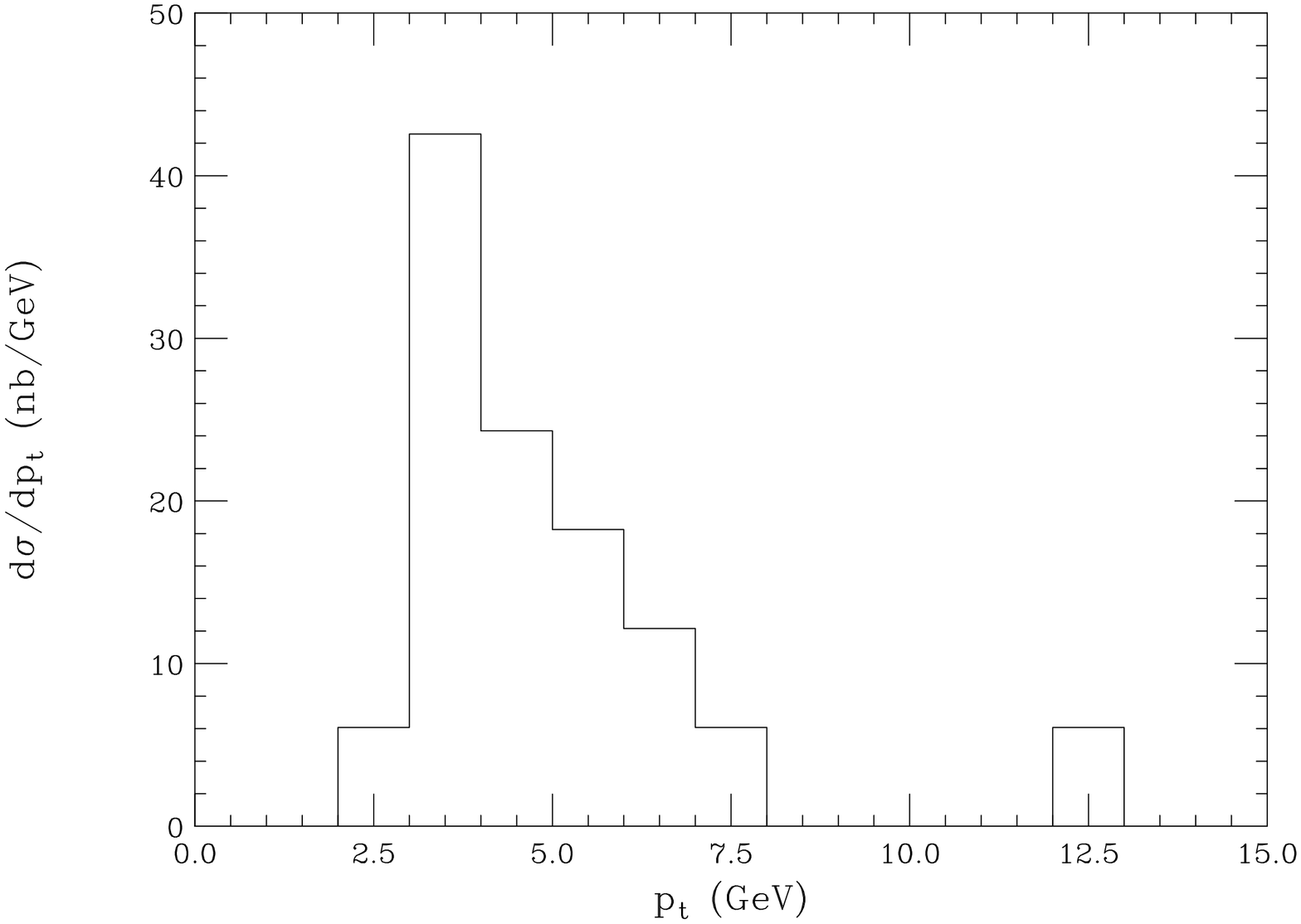,angle=90,width=0.8\textwidth}
\caption{Predicted \thetac\ transverse momentum distribution
in $p\bar p$ collisions at $\sqrt{S} = 2$ TeV.
\label{fig:thetac_pt_tev}} }

The $pD^*$ mass distributions before and after a cut on separation
are shown in fig.~\ref{fig:m_dist_tev}, corresponding to
figs.~\ref{fig:m_dist} and \ref{fig:m_dist_zdk}.
The predicted $\Theta_c$ transverse momentum distribution is shown in
fig.~\ref{fig:thetac_pt_tev}.
The number of pairs
in the coalescence region $0<\Delta x<2$ fm, $3050<m<3150$ MeV is 19.
Taking into account coalescence factor $F_{\rm co}\simeq 10$, 
this corresponds to 
190 \thetac\ events, i.e. to 
a \thetac\ production cross section of 285 nb.
Since the integrated luminosity already collected in
Run II is of the order of 200 pb$^{-1}$, this corresponds to
production of some 57 million \thetac's.  Even with an overall
detection efficiency of only a few per mille, this would seem to provide
an unmissable signal.

\section{Summary and conclusions}
We constructed a simple coalescence model for the production of the charmed
pentaquark reported by the H1 experiment. 
This model can also serve as a template for estimating resonance and 
bound state production cross sections in a wide range of hard processes.

In the specific case discussed here pentaquark formation is assumed
to occur when $D^* p$ are in close proximity in both momentum and
coordinate space.  Comparing the HERWIG-generated $D^* p$ distribution 
with the signal reported by H1, we determine the coalescence-enhancement
factor $F_{\rm co}\sim 10$. This is then applied to estimate the number of 
$\Theta_c$ events in $e^+e^-$ annihilation at LEP and $p \bar p$ collisions
at the Tevatron. For each of the four LEP experiments the model then
predicts between 25 and 40 H1-like $\Theta_c$ events. For the Tevatron
a signal of many thousands of events would have been expected. 

Since both LEP and Tevatron experiments reported null results, our analysis
implies that the either the H1 signal is spurious and 
due to an unknown systematic effect,
or alternatively that it corresponds to a real resonance, whose 
production mechanism in DIS is 
substantially different from the production mechanism in $e^+e^-$ and the
Tevatron. Yet another possibility is that either the theoretical or
experimental analysis is missing an essential ingredient.

The value of $F_{\rm co}\sim 10$ required to account for the H1 data
is surprisingly large. In the event that further analysis of the data
implies a smaller signal, or an upper limit, the value of
$F_{\rm co}$ can be scaled down to obtain the correspondingly
reduced predictions for $e^+e^-$ and $p\bar p$. It should be
noted that only the overall normalization of the predicted
distributions and not their shapes will be affected by this
procedure.
As remarked in sect.~2.2, one possible effect that would lead to
a reduced estimate of $F_{\rm co}$ is a higher proton yield in
charm production in DIS, relative to the \HW\ prediction.

 \acknowledgments

The research of one of us (M.K.) was supported in part by
a visiting fellowship grant from the Royal Society, by the
United States-Israel Binational Science Foundation (BSF), Jerusalem
and by
Israel Science Foundation administered by the Israel
Academy of Sciences and Humanities.
We would like to thank 
Yeon Sei Chung,
Karin Daum,
Leonin Gladilin,
Peter Hansen,
Uri Karshon
and
Terry Sloan 
for useful discussions of the experimental data.

\end{document}